\documentclass[reprint,aps,twoside,prd,superscriptaddress]{revtex4-2}
\pdfoutput=1 
\usepackage{graphicx}
\usepackage[T1]{fontenc}
\usepackage{amssymb}
\usepackage{amsmath}
\usepackage{tensor,mathtools}
\usepackage{dcolumn}
\usepackage{cancel} 
\usepackage{physics}

\usepackage{enumerate}

\usepackage[version=4]{mhchem}

\usepackage{siunitx} 
\DeclareSIUnit\year{yr}
\DeclareSIUnit \parsec {pc}
\usepackage{comment}

\usepackage{url}

\usepackage{aasmacros} 

\newcommand\numberthis{\addtocounter{equation}{1}\tag{\theequation}}

\usepackage[colorlinks]{hyperref}
\usepackage[usenames,dvipsnames]{color}
\hypersetup{
     breaklinks=true,
    pdfstartview={FitH},    
    colorlinks=true,       
    linkcolor=blue,          
    citecolor=red,        
    filecolor=magenta,      
    urlcolor=blue,           
    anchorcolor=green,      
    linktocpage=true
}

\newcommand{\afffias}{Frankfurt Institute for Advanced Studies (FIAS), Ruth-Moufang-Str.~1, 60438 Frankfurt am Main, Germany}
\newcommand{\affgoethe}{Physics Department, Goethe University, Max-von-Laue-Str.~1, 60438 Frankfurt am Main, Germany}

\date{\today}

\begin{document}

\title{Torsional dark energy in quadratic gauge gravity}
\author{A.~van de Venn}
\email{venn@fias.uni-frankfurt.de}
\affiliation{\afffias}
\affiliation{\affgoethe}
\author{D.~Vasak}
\email{vasak@fias.uni-frankfurt.de}
\affiliation{\afffias}
\author{J.~Kirsch}
\email{jkirsch@fias.uni-frankfurt.de}
\affiliation{\afffias}
\author{J.~Struckmeier}
\email{struckmeier@fias.uni-frankfurt.de}
\affiliation{\afffias}
\affiliation{\affgoethe}

\begin{abstract}
The Covariant Canonical Gauge theory of Gravity (CCGG) is 
a gauge field formulation of gravity which a priori includes
non-metricity and torsion. It extends the Lagrangian of Einstein's theory
of general relativity by terms at least quadratic in the
Riemann-Cartan tensor.
This paper investigates the implications of metric
compatible CCGG on cosmological scales. 
For a totally
anti-symmetric torsion tensor
we derive the resulting equations of
motion in a Friedmann-Lemaître-Robertson-Walker (FLRW) Universe. 
In the limit of a 
vanishing quadratic
Riemann-Cartan term, the arising modifications of the Friedmann 
equations are shown to be equivalent to spatial curvature.
Furthermore, the modified Friedmann equations are investigated
in detail in the early and late times of the Universe's
history. It is demonstrated that in addition to the standard
$\Lambda$CDM behaviour of the scale factor, there exist 
novel time dependencies, emerging due to the presence of 
torsion and the quadratic 
Riemann-Cartan term. Finally, at late times, we present how the
accelerated expansion of the Universe can be understood as a 
geometric effect of spacetime through torsion,
rendering the introduction of a 
cosmological constant redundant. In such a scenario
it is possible to compute an expected value for the parameters
of the postulated gravitational Hamiltonian/Lagrangian and
to provide a lower bound on the vacuum energy of matter. 
\end{abstract}

\maketitle

\section{Introduction} 
Mysterious components such as dark energy (DE) caused
Einstein's theory of general relativity (GR) to become
a incomplete and thus unsatisfactory explanation of gravity on cosmological scales.
As a result, there has been a plethora of attempts to modify GR, for 
example, by introducing additional degrees of freedom in the
form of scalar fields \cite{Galileon,Cov_Galileon,Horndeski} or vector fields 
\cite{Gen_Proca,Mass_Vector,Cosmic_Acc,Gen_Proca2}.

In this work however, we continue the path taken in
\cite{CCGG_Path1,CCGG_Path2,Cov_Ham,Struckid} which is to use the theory of
canonical transformations within the De Donder-Weyl formalism \cite{DeDonder,Weyl}
and the ideas of gauge theory
to arrive at a more general theory of gravity called
Covariant Canonical Gauge theory of Gravity (CCGG).
CCGG is based on merely four postulates \cite{CCGG_Ferm},
namely:
\begin{enumerate}[(i)]
    \item Hamilton's Principle
    \item Non-degeneracy of the total and gravitational Lagrangian
    \item Diffeomorphism invariance
    \item Equivalence Principle
\end{enumerate}

Employing the Palatini approach \cite{Palatini} of treating the
metric and the connection independently, these postulates,
together with the formalism of canonical transformations,
have been shown to result in the so-called
CCGG equations \cite{CCGG_Path1}
which generalise Einstein's field equation.
In particular, non-metricity and torsion are not a priori 
neglected within this approach.

As usual for gauge theories, the Lagrangian, or
respectively
the Hamiltonian, of the free fields cannot be determined and
has to be postulated. Non-degeneracy
restricts the choice of the free gravitational Hamiltonian
$\mathcal{H}_{\text{gr}}$ however
and forces us \cite{Benisty2018} to include at least a term quadratic in the 
Riemann-Cartan tensor. 
The ansatz for $\mathcal{H}_{\text{gr}}$ used in CCGG extends the Einstein theory by a trace-free Kretschmann term.
This has been shown to be consistent
with low-redshift data \cite{Low-red}.

The goal of this paper is to study the implications 
of the CCGG equations on cosmological scales,
setting forth earlier investigations 
\cite{CCGG_Cosm, CCGG_DE}.
In \autoref{sect_setup} a brief review of the underlying theory and the resulting 
CCGG field equation is presented.
In order to make contact with standard $\Lambda$CDM results, a
constraint on the torsion tensor and energy-momentum conservation is proposed in
\autoref{sect_Constr}.
A specific ansatz, satisfying this constraint, is then
presented and the resulting equations of motion are derived
in \autoref{temporal} with help of the xAct-package in
\textsc{mathematica} \cite{xAct}. Thereafter a brief
discussion is given in
\autoref{EC_lim} on the
limiting case in which the Kretschmann term vanishes. The key portion of this work is 
\autoref{CCGG_cosm}, where the asymptotic
equations of motion are
investigated in the early and late epochs of the
Universe.
Lastly, a summary and our conclusions complete this paper 
in \autoref{Concl}.

Throughout this paper we employ natural units, in which
$\hbar=c=1$.
Furthermore, the Misner-Thorne-Wheeler convention $(+++)$ 
\cite{Misner1973} is used.

\section{Setup}
\label{sect_setup}

The gauging process of CCGG has been worked out in 
detail within the framework of De Donder-Weyl theory in \cite{CCGG_Ferm} and results in the action integral
\begin{align}
S&=\int\mathrm{d}^4x \, \tilde{\mathcal{L}}\nonumber\\&= \int\mathrm{d}^4x\Biggl[\frac{1}{2}\tensor{\tilde{k}}{_i^{\mu\nu}}\Biggl(\pdv{\tensor{e}{^i_\mu}}{\tensor{x}{^\nu}}-\pdv{\tensor{e}{^i_\nu}}{\tensor{x}{^\mu}}
+\tensor{\omega}{^i_{j\nu}}\,\tensor{e}{^j_\mu}-\tensor{\omega}{^i_{j\mu}}\,\tensor{e}{^j_\nu} \Biggr)\nonumber\\
&+\frac{1}{2}\tensor{\tilde{q}}{_i^{j\mu\nu}}
\Biggl(\pdv{\tensor{\omega}{^i_{j\mu}}}{\tensor{x}{^\nu}}
-\pdv{\tensor{\omega}{^i_{j\nu}}}{\tensor{x}{^\mu}}
+\tensor{\omega}{^i_{n\nu}}\,\tensor{\omega}{^n_{j\mu}}
-\tensor{\omega}{^i_{n\mu}}\,\tensor{\omega}{^n_{j\nu}}\Biggr)\nonumber\\&-\mathcal{\Tilde{H}}_{\text{gr}}+\mathcal{\Tilde{L}}_{\text{matter}}\Biggr].\label{act_int}
\end{align}
Greek indices denote components with respect to a holonomic
basis whereas Latin indices refer to a non-holonomic basis.
The tetrads $\tensor{e}{^i_\alpha}$ translate between the
holonomic and non-holonomic bases. In this regard we have
$\tensor{g}{_{\mu\nu}} = \tensor{e}{^i_\mu}\tensor{e}{^j_\nu}
\tensor{\eta}{_{i j}}$, where $\tensor{\eta}{_{i j}}$ is the 
Minkowski metric.
A tilde always denotes a tensor density. Thus the ``momentum tensor density'' is given by
\begin{equation}
    \tensor{\Tilde{q}}{_{l}^{
    m\alpha\beta}}\equiv \varepsilon\,\tensor{q}{_{l}^{
    m\alpha\beta}} \coloneqq \pdv{\mathcal{\Tilde{L}}}{(\tensor{\partial}{_\beta}\tensor{\omega}{^l_{m\alpha}})},
\end{equation}
and denotes the canonical conjugate of the spin connection
$\tensor{\omega}{^l_{m\alpha}}$. Further, the conjugate of the
tetrads is 
\begin{equation}
    \tensor{\Tilde{k}}{_{l}^{
    \alpha\beta}} \equiv \varepsilon\,\tensor{k}{_{l}^{
    \alpha\beta}} \coloneqq \pdv{\mathcal{\Tilde{L}}}{(\tensor{\partial}{_\beta}\tensor{e}{^l_{\alpha}})}.
\end{equation}
Here $\mathcal{\Tilde{L}}$ denotes the Lagrangian density
of the whole system that is obtained by a complete Legendre 
transformation of the respective total Hamiltonian density.
Information regarding the matter content is incorporated
within $\mathcal{\Tilde{L}}_{\text{matter}}$.
The relevant Jacobi determinant in tetrad
formalism is ${\varepsilon \coloneqq \det(\tensor{e}{^i_\alpha}) \equiv \sqrt{-g}}$, where $g \coloneqq \det(\tensor{g}{_{\mu\nu}})$.

Varying \eqref{act_int} with respect to the dynamical fields
provides us with the set of canonical equations of motion
as in \cite{CCGG_Ferm}. These hold a priori for any arbitrary
choice of torsion and non-metricity tensors. In more detail,
our choice of the free gravity De Donder-Weyl Hamiltonian density $\mathcal{\Tilde{H}}_{\text{gr}}$ entails, amongst others, information
about the appearance of torsion and non-metricity.
One can show that the
ansatz \cite{CCGG_DE} for the free gravity De Donder-Weyl Hamiltonian density
\begin{align*} \label{def:Hamiltonian}
    \mathcal{\Tilde{H}}_{\text{gr}} = &
    -\frac{\tensor{\Tilde{q}}{_{l}^{
    m\alpha\beta}}\tensor{\Tilde{q}}{_{m}^{
    l\xi\lambda}}}{4 g_1 \varepsilon}\tensor{\eta}{_{k n}}
    \tensor{\eta}{_{i j}}\tensor{e}{^k_\alpha}\tensor{e}{^n_\xi}\tensor{e}{^i_\beta}\tensor{e}{^j_\lambda}
    \\
    &+ g_2\tensor{\Tilde{q}}{_{l}^{
    m\alpha\beta}}\tensor{\eta}{_{m n}}
    \tensor{e}{^l_\alpha}\tensor{e}{^n_\beta}
    \\
    &- \frac{\tensor{\Tilde{k}}{_{l}^{
    \alpha\beta}}\tensor{\Tilde{k}}{_{m}^{
    \xi\lambda}}}{2g_3 \varepsilon}\tensor{\eta}{^{l m}}
    \tensor{\eta}{_{k n}}\tensor{\eta}{_{i j}}
    \tensor{e}{^k_\alpha}\tensor{e}{^n_\xi}
    \tensor{e}{^i_\beta}\tensor{e}{^j_\lambda}
    - g_4 \varepsilon\numberthis
\end{align*}
induces the following CCGG field equation
\cite{CCGG_Ferm,CCGG_Path1,CCGG_Cosm,CCGG_DE}:
\begin{equation}
	-g_1\tensor{Q}{^\mu^\nu}+\frac{1}{8\pi
		G}\left(\tensor{G}{^\mu^\nu}+\tensor{g}{^\mu^\nu}\Lambda_0\right) 
    +2g_3\tensor{W}{^{\mu\nu}}=
	\tensor{T}{^{(\mu\nu)}},
	\label{CCGG}
\end{equation}
where $\tensor{g}{_{\mu\nu}}$ is the metric, and $g_i$ are fundamental coupling constants with the dimensions $[g_1] =1$, $[g_2] =L^{-2}$, ${[g_3] = L^{-2}}$ and $[g_4] = L^{-4}$.
The choice \eqref{def:Hamiltonian} ensures a vanishing
non-metricity (chosen here for simplicity) by requesting
$\tensor{q}{^{(l m)\alpha\beta}} = 0$ but a non-vanishing, a priori arbitrary, torsion.

We defined the symmetric and trace-free Kretschmann term
\begin{equation}
	\tensor{Q}{^\mu^\nu}\coloneqq
	\tensor{R}{^{\alpha\beta\gamma\mu}}\tensor{R}{_{\alpha\beta\gamma}
	^\nu}-\frac{1}{4}\tensor{g}{^{\mu\nu}}\tensor{R}{^{\alpha\beta\gamma
		\lambda}}\tensor{R}{_{\alpha\beta\gamma\lambda}},
\end{equation}
with the Riemann-Cartan tensor
\begin{equation}
	\tensor{R}{^\alpha_{\beta\mu\nu}} \coloneqq
	\tensor{\partial}{_\mu}\tensor{\Gamma}{^\alpha_{\beta\nu}}-
	\tensor{\partial}{_\nu}\tensor{\Gamma}{^\alpha_{\beta\mu}}+
	\tensor{\Gamma}{^\alpha_{\lambda\mu}}
	\tensor{\Gamma}{^\lambda_{\beta\nu}}-
	\tensor{\Gamma}{^\alpha_{\lambda\nu}}
	\tensor{\Gamma}{^\lambda_{\beta\mu}},
\end{equation}
and
\begin{equation}
    \tensor{W}{^\mu^\nu}\coloneqq \tensor{S}{^{\alpha\beta\mu}}
    \tensor{S}{_{\alpha\beta}^\nu}-\frac{1}{2}
    \tensor{S}{^{\mu\alpha\beta}}\tensor{S}{^{\nu}_{\alpha\beta}} - \frac{1}{4}\tensor{g}{^{\mu\nu}}
    \tensor{S}{^{\alpha\beta\gamma}}
    \tensor{S}{_{\alpha\beta\gamma}}
    \label{quadr_tors}
\end{equation}
with the torsion tensor 
\begin{equation}
	\tensor{S}{^\lambda_{\mu\nu}}\coloneqq
	\frac{1}{2}\left(\tensor{\Gamma}{^\lambda_{\mu\nu}}-
	\tensor{\Gamma}{^\lambda_{\nu\mu}}\right).
\end{equation}
The affine connection $\Gamma$ can thereby be decomposed as
\begin{equation}
	\tensor{\Gamma}{^\alpha_{\mu\nu}}=
	\tensor{\bar{\Gamma}}{^\alpha_{\mu\nu}} + 
	\tensor{K}{^\alpha_{\mu\nu}},
	\label{conn}
\end{equation}
where $\bar{\Gamma}$ is the Levi-Civita connection, and 
$\tensor{K}{^\alpha_{\mu\nu}}$ the contortion tensor.
(Henceforth any object equipped with a bar shall denote a
quantity based on $\bar{\Gamma}$.)
The contortion tensor is defined as
\begin{equation}
	\tensor{K}{_{\lambda\mu\nu}} \coloneqq 
	\tensor{S}{_{\lambda\mu\nu}}-\tensor{S}{_{\mu\lambda\nu}}+
	\tensor{S}{_{\nu\mu\lambda}}.
\end{equation}
The constant $\Lambda_0$ arising in Eq.~\eqref{CCGG} is related to the parameters $g_1,g_2$, governing the strength of the quadratic and linear terms in the Hamiltonian~\eqref{def:Hamiltonian}, and $g_4$ representing the vacuum energy of matter, as 
\begin{align*}
    g_1 g_2 &= -\frac{1}{16\pi G}\\
    6g_1 g_2^2 + g_4 &= \frac{\Lambda_0}{8\pi G}.
    \numberthis
    \label{parameters}
\end{align*}
Obviously $\Lambda_0$  can be identified with the cosmological constant.
The parameter $g_1$ in particular regulates the relative strength of the trace-free Kretschmann term ``deforming'' Einstein-Cartan gravity.

Care has to be taken with the Einstein tensor
$\tensor{G}{^\mu^\nu}$. Here it is defined to only include the symmetric part of the Ricci tensor
\begin{equation}
	\tensor{G}{^\mu^\nu}\coloneqq \tensor{R}{^{\left(\mu\nu\right)}}-
	\frac{1}{2}\tensor{g}{^\mu^\nu}R.
\end{equation}
Since torsion is present, $\tensor{R}{^{\mu\nu}}$ also has an anti-symmetric portion which,
together with the anti-symmetric portion of the canonical energy momentum tensor, $\tensor{T}{^{[\mu\nu]}}$, yields another set of equations \cite{CCGG_Path1, CCGG_Path2}
relating  torsion and the spin density of matter. 
However, for our purposes in this paper we can ignore this set of equations and focus solely on the symmetric equations \eqref{CCGG}. 

The gauging procedure further reveals that the energy-momentum
tensor of spacetime itself 
corresponds to the l.h.s. of \eqref{CCGG} up to a minus sign.
Upon defining
\begin{equation}
	\tensor{\Theta}{^{\mu\nu}} \coloneqq -g_1\tensor{Q}{^\mu^\nu}+\frac{1}{8\pi
		G}\left(\tensor{G}{^\mu^\nu}+\tensor{g}{^\mu^\nu}\Lambda_0\right)+ 2g_3 \tensor{W}{^{\mu\nu}}
\end{equation}
the CCGG equations \eqref{CCGG} are compactly written as
\begin{equation}
	\tensor{\Theta}{^{\mu\nu}} = \tensor{T}{^{(\mu\nu)}}.
	\label{CCGG2}
\end{equation}

\section{Constraints}
\label{sect_Constr}
It is straightforward to verify that  $\tensor{\nabla}{_\nu}\tensor{\Theta}{^{\mu\nu}} = 0$
is not an identity.
Thus in general, the energy-momentum conservation
$\tensor{\nabla}{_\nu}\tensor{T}{^{(\mu\nu)}} = 0$ is not 
necessarily satisfied either.
In order to determine the evolutionary behaviour of 
energy densities on cosmological scales, we therefore
have to ensure
\begin{equation}
    \tensor{\nabla}{_\nu}\tensor{\Theta}{^{\mu\nu}} = 
    \tensor{\nabla}{_\nu}\tensor{T}{^{(\mu\nu)}}.
    \label{new_EM_cons}
\end{equation}
However, \eqref{new_EM_cons} is difficult to solve in general. 
In order to make the computations analytically tractable we have to make a few simplifying assumptions.

For any symmetric tensor $\tensor{A}{^{\mu\nu}}$ we have that
\begin{equation}
    \tensor{\nabla}{_\nu}\tensor{A}{^{\mu\nu}} = 
    \tensor{\bar{\nabla}}{_\nu}\tensor{A}{^{\mu\nu}}+
	\tensor{K}{^\mu_{\alpha\nu}}\tensor{A}{^{\alpha\nu}}+
    \tensor{K}{^\nu_{\alpha\nu}}
	\tensor{A}{^{\mu\alpha}}.
	\label{div_symm}
\end{equation}
Closer inspection of the second term on the r.h.s. reveals
$\tensor{K}{^\mu_{\alpha\nu}}\tensor{A}{^{\alpha\nu}}=
\tensor{K}{^\mu_{\left(\alpha\nu\right)}}
\tensor{A}{^{\alpha\nu}}$, due to the symmetry of $\tensor{A}{^{\mu\nu}}$, with 
\begin{align*}
\tensor{K}{^\mu_{\left(\alpha\nu\right)}} = \frac{1}{2}\Bigl(
&\tensor{S}{^\mu_\alpha_\nu}-\tensor{S}{_\alpha^\mu_\nu}
+\tensor{S}{_\nu_\alpha^\mu}\\&+ \tensor{S}{^\mu_\nu_\alpha}
-\tensor{S}{_\nu^\mu_\alpha}+\tensor{S}
{_\alpha_\nu^\mu}\Bigr).\numberthis
\end{align*}
Since $\tensor{S}{^\lambda_\mu_\nu} =
-\tensor{S}{^\lambda_\nu_\mu}$ it follows that
\begin{equation}
	\tensor{K}{^\mu_{\left(\alpha\nu\right)}} = 
	2\tensor{S}{_{\left(\alpha\nu\right)}^\mu}.
\end{equation}
For the third term in \eqref{div_symm} we find
\begin{equation}
    \tensor{K}{^\nu_{\alpha\nu}} = 2 \tensor{S}{^\nu_{\alpha\nu}}.
\end{equation}
If $\tensor{S}{_{\alpha\nu\mu}}$ is additionally
anti-symmetric in its
first two indices then $\tensor{S}{_{\left(\alpha\nu\right)}_\mu}=0$
and $\tensor{S}{^\nu_{\alpha\nu}} = 0$.
Therefore we conclude that for a totally anti-symmetric torsion tensor the divergence of a symmetric $(2,0)$-tensor
is equal to the divergence based solely on the Levi-Civita connection 
$\tensor{\nabla}{_\nu}\tensor{A}{^{\mu\nu}} = \tensor{\bar{\nabla}}{_\nu}\tensor{A}{^{\mu\nu}}$.
    By assuming a totally anti-symmetric torsion tensor henceforth, relation \eqref{new_EM_cons} becomes
\begin{equation}
    \tensor{\bar{\nabla}}{_\nu}\tensor{\Theta}{^{\mu\nu}} = 
    \tensor{\bar{\nabla}}{_\nu}\tensor{T}{^{(\mu\nu)}}
\end{equation}
due to the symmetry of $\tensor{\Theta}{^{\mu\nu}}$.

In the following we wish to apply this ansatz in a cosmological setting based on the Friedmann-Lemaître-Robertson-Walker (FLRW) metric, and align the analysis with the assumptions of the standard model as far as possible. 
Hence we require in addition that $\tensor{\bar{\nabla}}{_\nu}\tensor{T}{^{(\mu\nu)}} = 0$ holds, 
and thus by consistency also $\tensor{\bar{\nabla}}{_\nu}\tensor{\Theta}{^{\mu\nu}} = 0$.
This restriction allows to employ the scaling behaviour of the individual energy densities as known from standard cosmology. The second equation, $\tensor{\bar{\nabla}}{_\nu}\tensor{\Theta}{^{\mu\nu}} = 0$, must be verified case by case, though. 
Fortunately this turns out to be rather easy here.

\section{Modified Friedmann equations}
\label{temporal}
As discussed in the previous section we choose a totally anti-symmetric ansatz for torsion,
\begin{equation}
    \tensor{S}{_{\alpha\mu\nu}} = \tensor{
    \epsilon}{_{\alpha\mu\nu\sigma}}\tensor{s}{^\sigma},
    \label{antisymm_tors}
\end{equation}
where $\tensor{\epsilon}{_{\alpha\mu\nu\sigma}}$ is the
Levi-Civita tensor and ${\tensor{s}{^\mu} = 
\left(s_0(t),0,0,0 \right)}$ a temporal axial four-vector. This 
ansatz has already been successfully used in 
\cite{dark_side_tors}. 
Since our underlying theoretical framework poses no restrictions on the form of the torsion tensor, the only 
caveat is that we require \eqref{antisymm_tors} to be 
consistent with the Cosmological Principle, i.e., a spatially homogeneous and isotropic Universe.
In fact, the allowed values for the torsion tensor, given a
FLRW-metric, have already been worked out in full generality
by \cite{Tsamparlis} and have been employed prominently by 
\cite{Iosifidis1,Iosifidis2,Fabbri}.
Our choice \eqref{antisymm_tors} is now seen to be in
correspondence with the
Cosmological Principle as it is a subset of the generally
allowed values for the torsion tensor in a FLRW-Universe \cite{Tsamparlis}.
Moreover, this choice ensures that the auto-parallel and geodesic trajectories of test point particles in curved geometry coincide.
The FLRW-metric in spherical coordinates is
\begin{equation}
    \mathrm{d}s^2 = -\mathrm{d}t^2 + a^2(t)\left(\frac{\mathrm{d}r^2}{1-kr^2} + 
    r^2\mathrm{d}\Omega^2\right),
\end{equation}
where $\mathrm{d}\Omega^2 = 
\mathrm{d}\theta^2 + \sin^2(\theta)\mathrm{d}\phi^2$. 
As usual, $a(t)$ is the scale factor and $k$ denotes the spatial curvature. 
Matter is modelled by a set of non-interacting perfect fluids with the energy-momentum tensor
\begin{equation}
    \tensor{T}{^{\mu}^{\nu}} = \left(\rho + p\right)
    \tensor{u}{^\mu}\tensor{u}{^\nu} +
    p\tensor{g}{^{\mu\nu}}.
\end{equation}
The total energy density $\rho$ and pressure $p$ are the sums of the individual constituents, which in this case are radiation and matter (baryonic and non-baryonic).
The vector field $\tensor{u}{^\mu} = (1,0,0,0)^\intercal$
denotes the 4-velocity of the fluid.
Since $\tensor{K}{_{\alpha\mu\nu}} \equiv \tensor{S}{_{\alpha\mu\nu}}$ for a totally anti-symmetric torsion tensor, we are now in the position to express the connection~\eqref{conn} and thus the components of the CCGG field equations explicitly.

Eq.~\eqref{CCGG2} reduces to only two independent components which give modified versions of the standard Friedmann equations:
\begin{multline}
 3 g_1 \Biggl[\frac{k^2}{a^4} + \frac{2k}{a^2}\left(H^2 - s_0^2\right)-H^2 \left(2 \dot{H}+5 s_0^2\right)\\-\dot{H}^2+2 H s_0 \dot{s}_0+\dot{s}_0^2+s_0^4\Biggr] +\rho_m + \rho_r \\+\frac{-3 H^2+\Lambda_0+3 s_0^2\left(1-8\pi G g_3\right)}{8 \pi  G} - \frac{3k}{8\pi G a^2}=0
 \label{Fried1}
\end{multline}
and
\begin{multline}
 g_1 \Biggl[\frac{k^2}{a^4}+\frac{2k}{a^2}\left(H^2 - s_0^2\right)-H^2 \left(2 \dot{H}+5 s_0^2\right)\\-\dot{H}^2+2 H s_0 \dot{s}_0+\dot{s}_0^2+s_0^4\Biggr]
+p_r \\+ \frac{3H^2 + 2\dot{H} - \Lambda_0 - s_0^2\left(1-8\pi G g_3\right)}{8\pi G}
+\frac{k}{8 \pi G a^2}=0.
\label{Fried2}
\end{multline}
Here we used $\rho = \rho_r + \rho_m$ and assumed the usual
equation of state $p_i=w_i\,\rho_i$ with $w_m=0$ for matter and $w_r=1/3$ for radiation. The Hubble function is defined as
${H \coloneqq \dot{a}/a}$.

The trace of \eqref{CCGG2} is found to be
\begin{align}
     \rho_m &+ \frac{-3 \dot{H}-6 H^2+2 \Lambda_0+3 s_0^2\left(1-8\pi G g_3\right)}{4 \pi  G} \nonumber\\
     &- \frac{3k}{4\pi G a^2}=0.
    \label{trace}
\end{align}
Note that there is no $g_1$ term appearing in \eqref{trace}, i.e., there is no
contribution from the quadratic ``radiation like'' Riemann term
since $\tensor{Q}{^\mu_\mu}=0$.

\section{Einstein-Cartan Limit}
\label{EC_lim}
Let us first consider the limit $g_1 \to 0$. This corresponds
to a modification of standard GR based solely on the introduction
of a non-vanishing torsion tensor. In other words, this case
corresponds to Einstein-Cartan theory with a special ansatz
for $\tensor{S}{_{\alpha\mu\nu}}$. However, our ansatz
for torsion differs from the one used, e.g., in \cite{Kranas}.
Therefore we expect to find a different behaviour 
in this regime.

The Friedmann equations \eqref{Fried1} and \eqref{Fried2} reduce
to 
\begin{equation}
    H^2 = \frac{8\pi G}{3}\rho + \frac{\Lambda_0}{3}
    -\frac{k}{a^2} + s_0^2\left(1-8\pi G g_3\right)
    \label{Fried1_EC}
\end{equation}
and 
\begin{equation}
    \frac{\ddot{a}}{a} = -\frac{4\pi G}{3}\left(\rho + 3p_r\right)+ \frac{\Lambda_0}{3}.
    \label{Fried2_EC}
\end{equation}
Furthermore, \eqref{Fried1} and \eqref{Fried2} in the limit of $g_1 \to 0$ reveal that the remaining contribution of the torsion may be absorbed into the energy-momentum tensor of a perfect fluid with energy density
\begin{equation}
    \rho_s = \frac{3s_0^2\left(1-8\pi G g_3\right)}{8\pi G}
\end{equation}
and pressure
\begin{equation}
    p_s = -\frac{s_0^2\left(1-8\pi G g_3\right)}{8\pi G}.
\end{equation}
Thus torsion admits the equation of state with 
${w_s = -1/3}$,
same as for the spatial curvature $k$.
Since we were able to define 
an appropriate energy density and pressure for the torsion
terms, conservation of the torsion contribution is
automatically ensured
via the standard energy-momentum conservation
${\tensor{\bar{\nabla}}{_\nu}\tensor{T}{^{(\mu\nu)}} = 0}$.
Hence we know that the energy density associated to $s_0$ scales as $\rho_s \propto a^{-2}$ and thus $s_0 \propto a^{-1}$. 

The l.h.s. of the CCGG equations in this case just 
reduces to the Einstein tensor (based solely on
the Levi-Civita connection) together with the cosmological
constant term
and hence satisfies
${\tensor{\bar{\nabla}}{_\nu}\left(\tensor{G}{^{\mu\nu}}
+ \tensor{g}{^{\mu\nu}}\Lambda_0\right) = 0}$ 
due to the Bianchi identities and metricity.
This therefore ensures that 
\begin{equation}
    \tensor{\bar{\nabla}}{_\nu}\tensor{\Theta}{^{\mu\nu}}=0
    \label{cons_lhs}
\end{equation}
is consistently satisfied.

As $s_0$ enters into the connection \eqref{conn} linearly,
it has to admit real values. 
By setting $s_0\sqrt{1-8\pi G g_3} = b/a$ for some 
 $b$ with $[b] = \SI{}{L}^{-1}$, 
equation \eqref{Fried1_EC} becomes
\begin{equation}
    H^2 = \frac{8\pi G}{3}\rho + \frac{\Lambda_0}{3}
    -\frac{k-b^2}{a^2}.
\end{equation}
Obviously, the contribution of torsion counteracts the spatial
curvature \cite{CCGG_Cosm,CCGG_DE}. 
Therefore it is possible to misinterpret in the standard 
$\Lambda$CDM model the geometry type as open, albeit
$k \geq 0$, as long as $k-b^2<0$, or flat with $k=b^2>0$.
A closed Universe still appears only with a positive $k$, however with the slightly stronger constraint $k > b^2$.

With density parameters defined by 
\begin{align*}
    \Omega_r &= \frac{8\pi G}{3 H^2}\rho_r\\
    \Omega_m &= \frac{8\pi G}{3 H^2}\rho_m\\
    \Omega_\Lambda &=\frac{\Lambda_0}{3H^2}\\
    \Omega_k &= -\frac{k}{a^2 H^2}\\
    \Omega_s &= \frac{s_0^2\left(1-8\pi G g_3\right)}{H^2}, \numberthis
    \label{dens_params}
\end{align*}
Eq.~\eqref{Fried1_EC} is equivalent to  
\begin{equation}
    1 = \Omega_r + \Omega_m + \Omega_\Lambda + \Omega_k + \Omega_s.
\end{equation}
Thus, torsion dominates if $s_0^2\left(1-8\pi G g_3\right) \to H^2$ and vanishes trivially for $s_0 \to 0$ or 
$g_3 \to 1/(8\pi G)$.

\begin{figure}
\includegraphics[width=8.6cm,height=5cm]{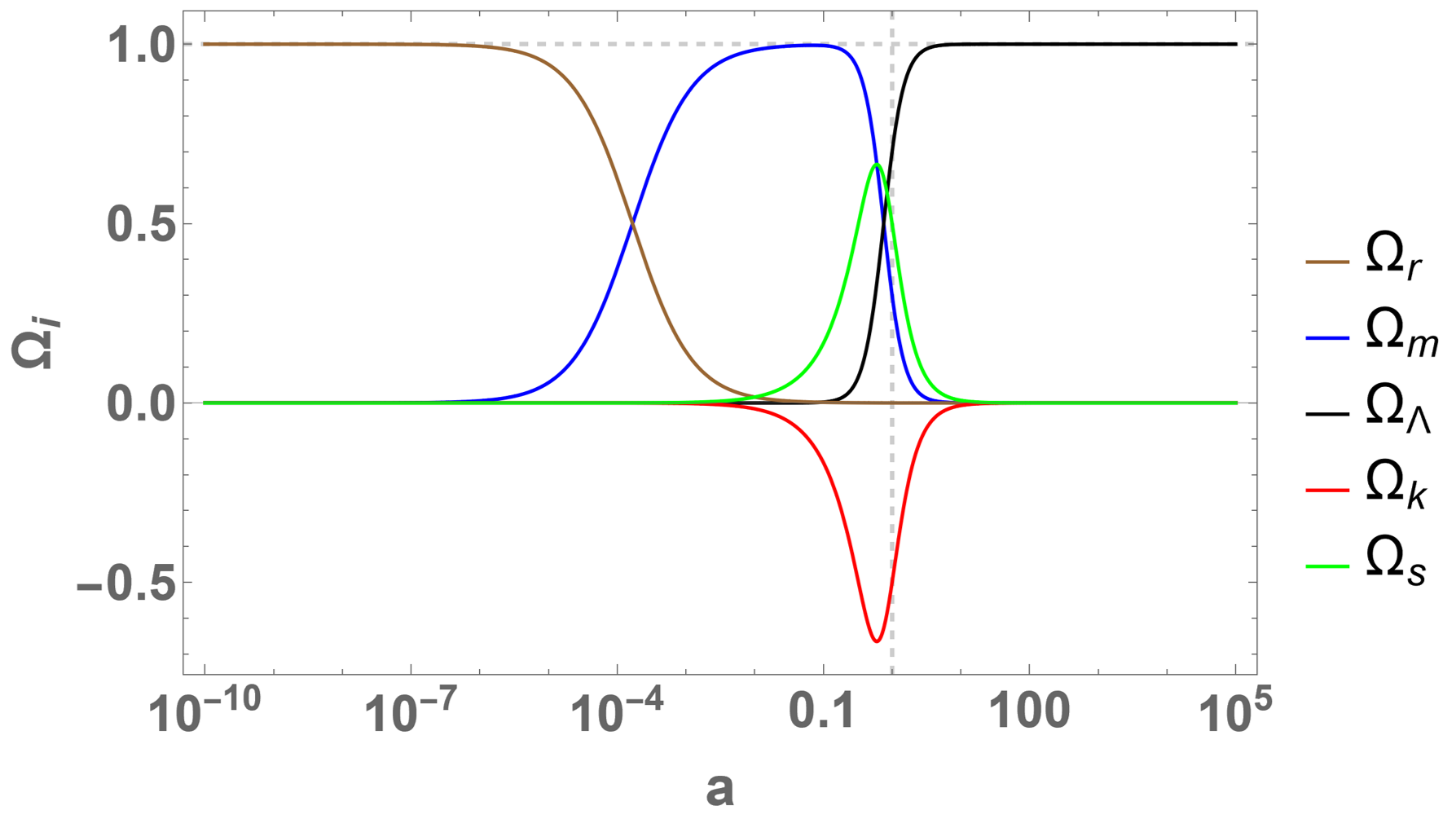}
\caption{Qualitative plot of the density parameters as a function of the scale factor with logarithmic scale in $a$. The chosen values for the density parameters today are 
$\Omega_{r,0} = 5\times 10^{-5}$, $\Omega_{m,0} = 0.3$, $\Omega_{\Lambda,0} = 0.7$ and $\Omega_{k,0} = -0.5$. The 
inferred value for the torsion density parameter today
is therefore $\Omega_{s,0} = -(\Omega_{k,0} + \Omega_{r,0})
= 0.49995$.}
\label{Dens_param}
\end{figure}

Let us choose $a(t=t_0) = 1$, where $t_0$ denotes today.
The Hubble constant is denoted by $H_0 := H(1)$. Then 
\begin{equation}
    \frac{H^2}{H_0^2} = \Omega_{r,0}\,a^{-4} + \Omega_{m,0}\,a^{-3}
    +  \Omega_{\Lambda,0} +  \Omega_{k,0}\,a^{-2} + 
     \Omega_{s,0}\,a^{-2},
\end{equation}
where $\Omega_{i,0}$ is the $i$-th density parameter
evaluated at ${t=t_0}$. Using 
\begin{equation}
    \Omega_i = \Omega_{i,0}\frac{H_0^2}{H^2}a^{-n_i}
\end{equation}
with $n_r = 4$, $n_m = 3$, $n_{\Lambda}=0$ and 
$n_k = n_s = 2$, we may then compute the behaviour of the
density parameters as a function of the scale factor for
given values of $\Omega_{i,0}$.
In Fig.\,\ref{Dens_param} these behaviours are shown for
${\Omega_{r,0} = 5\times 10^{-5}}$, $\Omega_{m,0} = 0.3$, $\Omega_{\Lambda,0} = 0.7$ and 
(for illustrative purposes) $\Omega_{k,0} = -0.5$.
As expected, the contribution of the torsion counteracts the contribution of the spatial curvature and radiation today, i.e.~$\Omega_{k,0}+\Omega_{r,0}=-\Omega_{s,0}$ in this case.

\section{Evolution of the universe}
\label{CCGG_cosm}
Let us now consider the general case $g_1 \neq 0$
and focus on investigating the asymptotic behaviour of the Friedmann equations \eqref{Fried1} and \eqref{trace}
(which is equivalent
to considering \eqref{Fried1} and \eqref{Fried2}).

\subsection{Radiation dominated epoch}
In the radiation dominated epoch (RDE) we choose the ansatz 
$a = \beta t^{\alpha}$ with
some constants $\alpha$ and $\beta$, where 
$[\beta] = \SI{}{L}^{-\alpha}$. Then 
$H = \alpha t^{-1}$ and 
${\dot{H} = -\alpha t^{-2}}$.
Assuming that the energy content of the Universe
is dominated by
radiation and relativistic particles,
we may neglect the cosmological constant and matter
energy density. 
For simplicity we further assume the spatial curvature to vanish, and leave the case with $k\ne 0$ to a forthcoming numerical study \cite{Tors_cosmol}. This allows us to find an analytic solution of Eq.~\eqref{trace}:
\begin{equation}
    s_0 = \pm \sqrt{\frac{\alpha\left(2\alpha - 1\right)}{1-8\pi G g_3}}\,\frac{1}{t}.
    \label{s0_early}
\end{equation}
We directly infer the special cases $\alpha = 0$ and 
$\alpha = 1/2$, which both yield a vanishing torsion
contribution. As it turns out, the remaining consistent 
solutions for $s_0$ will be such that 
$\alpha\left(2\alpha - 1\right) \geq 0$. Hence we need
to impose $g_3 < 1/(8\pi G)$
in order to obtain physical solutions.
Aforementioned approximations together with \eqref{s0_early} and ${\rho_r = \rho_{r,0}a^{-4}}$ transform the first Friedmann equation \eqref{Fried1} into
\begin{equation}
    \frac{\rho_{r,0}}{\beta^4t^{4\alpha}}+
    \frac{C_1(\alpha)}{t^2}-
    g_1\frac{C_2(\alpha)}{t^4} = 0.
    \label{Fried1_early}
\end{equation}
The coefficient
\begin{equation}
    C_1(\alpha) \coloneqq \frac{3\alpha\left(\alpha-1\right)}{8\pi G}
\end{equation}
contains the explicit contributions from the Einstein tensor
and the quadratic torsion tensor $\tensor{W}{^{\mu\nu}}$,
whereas 
\begin{align}
    C_2(\alpha) &\coloneqq \frac{3\alpha\left(2\alpha - 1\right)}{(1-8\pi G g_3)^2}\Bigl[(\alpha + 1)(3\alpha -1)
    \nonumber\\
    &\,\,\,\,\,- 
    8\pi G g_3(5\alpha^2 - 1) - 64\pi^2 G^2 g_3^2 \alpha\Bigr]
\end{align}
comprises the contributions from the quadratic Riemann-Cartan
term.
Naturally, $C_2$ implicitly carries the former contributions
via \eqref{s0_early}.

Equation \eqref{Fried1_early} has to hold as a polynomial 
equation for small $t$ with $t \neq 0$ and thus
the respective coefficients of all monomials have to be zero.
Due to the appearance of $\alpha$
in the exponent we have to distinguish between 
different cases.
The cases are
$\alpha = 1$, $\alpha = 1/2$ and all other real values, i.e., $\alpha \notin \{1,1/2\}$. However,
for $\alpha \notin \{1,1/2\}$ we know from \eqref{Fried1_early} that $C_1(\alpha) = 0$ and 
${C_2(\alpha) = 0}$ have to hold, which is only true for 
$\alpha = 0$. This implies $\rho_{r,0} = 0$,
which contradicts our assumption and hence we are left with
\begin{equation}
    \alpha \in \{1,\frac{1}{2}\}.
    \label{alphval}
\end{equation}
For $\alpha = 1/2$ we find $s_0 = 0$ and
$C_2(1/2) = 0$. Furthermore we need to have
\begin{equation}
    \frac{\rho_{r,0}}{\beta^4}+
    C_1\left(\frac{1}{2}\right) \stackrel{!}{=}0.
\end{equation}
But 
$C_1(1/2) = -3/32\pi G$ and thus
$\beta^4 = 32\pi G\rho_{r,0}/3$. Since only non-negative, real solutions for the scale factor are admissible we find
\begin{equation}
    a = \sqrt[4]{\frac{32\pi G}{3}\rho_{r,0}}\,\sqrt{t}.
\end{equation}
Note that this result corresponds exactly to the scale factor in the RDE of standard cosmology.
This is to be expected, since for $\alpha = 1/2$ the
contribution of the quadratic Riemann-Cartan term
and the torsion vanishes and hence we find ourselves in the
regime of Einstein's GR.

For the last case $\alpha = 1$ we have
\begin{equation}
    s_0 = \pm  \frac{1}{\sqrt{1-8\pi G g_3}}\,\frac{1}{t}
\end{equation}
and $C_1(1) = 0$. Thus we are left with 
\begin{equation}
    \frac{\rho_{r,0}}{\beta^4}-
    g_1 C_2\left(1\right) \stackrel{!}{=}0.
\end{equation}
With 
\begin{equation}
    C_2(1) = \frac{12 - 96\pi G g_3\left(1+ 2\pi G g_3\right)}{(1-8\pi G g_3)^2}
\end{equation}
we get $\beta^4 = \rho_{r,0}/(g_1 C_2(1))$
and therefore
\begin{equation}
    a = \sqrt[4]{\frac{\rho_{r,0}}{g_1 C_2(1)}}\,t,
    \label{scale_lin}
\end{equation}
which requires $g_1 C_2(1) > 0$ and $g_3 \neq (-1\pm\sqrt{2})/(4\pi G)$.
This solution yields an interesting time dependence of the
scale factor resulting in particular from the presence of the
quadratic Riemann-Cartan term and the torsion. The linear time
dependence is in correspondence with that of a Milne Universe.
However, it is achieved in a very different manner.
Namely, in contrast to a Milne Universe we did assume $k=0$
and $\rho_{r,0} \neq 0$. In addition we have a torsion contribution that decreases as the Universe expands.

To understand how the different ingredients of the CCGG
equations contribute to the total energy density, we define
the density parameter 
\begin{align*}
    \Omega_{\text{geo}} &= \frac{8\pi G g_1}{H^2} \Bigl[
    -H^2(2\dot{H}+5s_0^2)-\dot{H}^2\\
    &\qquad\qquad\quad+ 2s_0\dot{s_0}H
    + \dot{s_0}^2 + s_0^4\Bigr]. \numberthis
    \label{RDE_params}
\end{align*}
The parameters $\Omega_{r}$ and $\Omega_{s}$ that are relevant in this scenario are defined exactly in the
same manner as in the Einstein-Cartan limit 
\eqref{dens_params}. They incorporate the energy density of 
radiation and the energy density of torsional contributions 
from $\tensor{G}{^{\mu\nu}}$ and $\tensor{W}{^{\mu\nu}}$.
The novel $\Omega_{\text{geo}}$ on the other hand 
describes the energy density of the Kretschmann term
$\tensor{Q}{^{\mu\nu}}$. The 
first Friedmann eq. \eqref{Fried1}, in this RDE epoch,
thus reduces to
\begin{equation}
    1 = \Omega_{r} + \Omega_{s} + \Omega_{\text{geo}}.
\end{equation}
We further have 
\begin{align*}
    \Omega_{r} &= \Omega_{r,0}a^{-4}\frac{H_0^2}{H^2}\\
    \Omega_{s} &= \Omega_{s,0}a^{-2}\frac{H_0^2}{H^2}
    \numberthis
    \label{RDE_today}
\end{align*}
with 
\begin{equation}
    \frac{H_0^2}{H^2} = a^2 H_0 \sqrt{\frac{8\pi G g_1 C_2(1)}{3\Omega_{r,0}}}
    \label{H_0/H}
\end{equation}
for the case $a \propto t$. 

Now notice that our assumptions
for the RDE are only valid in the very early Universe. 
Nevertheless we still extrapolate the solution until the 
present time
by referencing today's values of the density parameters
\eqref{RDE_today} and the Hubble constant $H_0$.
This is obviously not consistent. We should rather  
normalise $a=1$ at a time in the very early Universe,
and reference the density parameters to that same time.
Unfortunately we neither know the exact value of the Hubble 
parameter at times in the very early Universe, nor do we
know the radiation density.
Thus, to limit the amount of unknown parameters, we stick
to the procedure above. The result will not provide us with
true numerical values, but give us merely a qualitative idea of the 
contributions of the different ingredients in the RDE.

\begin{figure}
\includegraphics[width=8.6cm,height=5cm]{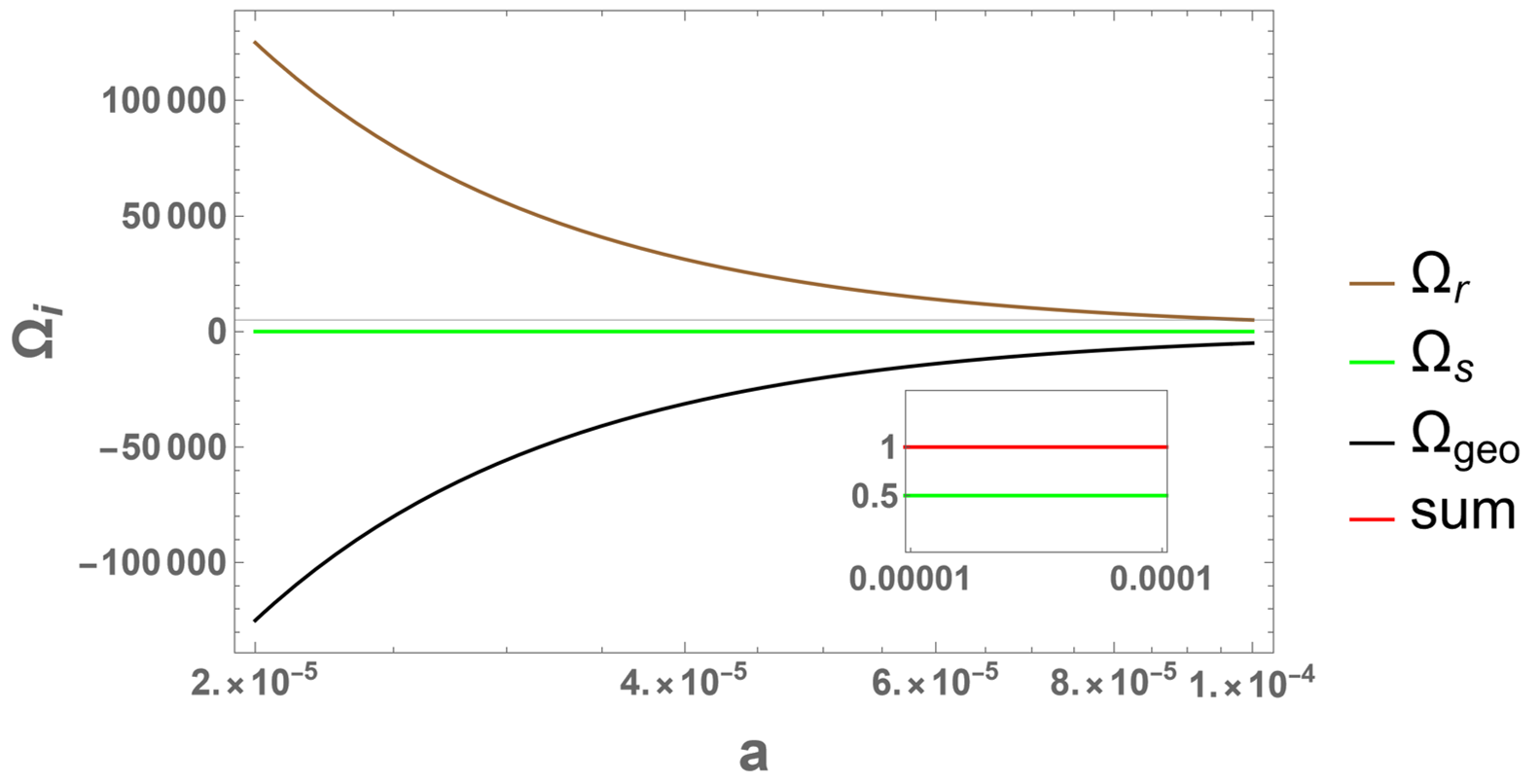}
\caption{Behaviour of the density parameters as a function
of the scale factor with logarithmic scale in $a$. The values
of the density parameters today were chosen to be
${\Omega_{r,0} = 5\times10^{-5}}$ and $\Omega_{s,0} = 0.5$.
Furthermore, $g_3 = 0.34\, M_p^2$ was taken, meaning $g_1 = 2.3\times 10^{115}$ via \eqref{g1RDE}.}
\label{RDE_dens}
\end{figure}

Evaluating now \eqref{H_0/H} at today yields
for a consistency relation between
different, still undetermined parameters:
\begin{equation}
    g_1 = \frac{3\Omega_{r,0}}{8\pi G H_0^2 C_2(1)}.
    \label{g1RDE}
\end{equation}
In particular, given $\Omega_{r,0}$ and $H_0$, the value of
$g_1$ is fixed if we provide a specific value for $g_3$.
For ${H_0 = \SI{70}{\kilo\metre\per\second\per\mega\parsec}}$,
$\Omega_{r,0} = 5\times10^{-5}$, $\Omega_{s,0} = 0.5$ and
$g_3 = 0.34\, M_p^2$, where 
$M_p \coloneqq 1/\sqrt{8\pi G} = \SI{2.44e18}{\giga\electronvolt} $ 
is the reduced Planck mass, the density parameters
are shown in Fig.\,\ref{RDE_dens}. 
Beside the negligible 
impact of $\Omega_s$ in the early times, we see in particular
the important contribution from the Kretschmann term
$\Omega_{\text{geo}}$ that counters the radiation energy
density. 
The contributions of 
$\Omega_{\text{geo}}$ increase with decreasing time/scale factor. 
This is expected since the Kretschmann term is
believed to yield high contributions in high density 
environments.

From \eqref{Fried1} and \eqref{Fried2} it is evident that
since the Kretschmann term is trace-free, then, if considered as a perfect
fluid, it admits an equation of state parameter $w = 1/3$, same
as that of radiation. 
Based on the qualitative discussion above
we argue that at sufficiently early times, $\Omega_{\text{geo}}$ must admit negative values and thus
behave like negative radiation, i.e., radiation with negative
energy density. Friedmann cosmologies with such negative 
energy densities have been investigated for instance in
\cite{Nemiroff,Akarsu}. At a certain point in time, depending on the 
exact value of $\Omega_s$, $\Omega_{\text{geo}}$ might have 
to switch its sign and become positive in order to satisfy the
Friedmann equation.
In general we thus conclude that the 
trace-free Kretschmann term is ghost-like, and its dynamics
can provide both, positive or negative energy density.

\medskip
Let us lastly investigate whether the consistency with
energy-momentum conservation, i.e., Eq.~\eqref{cons_lhs} is
fulfilled.
It is straightforward to see that with the given assumptions,
$\tensor{\bar{\nabla}}{_\nu}\tensor{\Theta}{^{\mu\nu}}=0$
reduces, for $\mu = 0$, to 
\begin{align} \label{early_consist}
    &3\alpha (\alpha-1)(2\alpha-1) \times \nonumber \\ 
    &\Bigl\{(1 - 8\pi G g_3)^2t^2
    - 16 \pi G g_1 \Bigl[
    (\alpha+1) (3\alpha-1) \nonumber \\
    &+8\pi G g_3\left(1-5\alpha^2 - 8\pi G g_3 \alpha\right)\Bigr]\Bigr\}=0.
\end{align}
The other three equations for $\mu \in \{1,2,3\}$
are trivial as in the Einstein-Cartan limit.
Indeed we see that \eqref{alphval} satisfies this consistency
check.
Equation \eqref{early_consist} reveals two more possible
values for $\alpha$, namely, the roots of the expression
inside the square bracket.
However, these roots are time dependent and hence
contradict our assumption of $\alpha$ being a constant.

\subsection{Dark energy dominated epoch}
Here we wish to address
the late time accelerated expansion of the
Universe and its relation to
torsion and the quadratic Riemann-Cartan term. Note that 
torsion alone is not able to account for an accelerated
expansion in the Einstein-Cartan limit. This is most easily 
seen in the second Friedmann equation \eqref{Fried2_EC} where
no torsion term appears and thus cannot alter $\ddot{a}$ in a
torsion dominated epoch.
On the other hand, as we will see, the full CCGG theory
allows for such a scenario. Thus the quadratic Riemann-Cartan
term is vital for the following steps.

\subsubsection{General setup}
An accelerated expansion is achieved if
\begin{equation}
    \ddot{a} > 0 \quad \text{and} \quad \dot{a}>0.
\end{equation}
Now let us assume that torsion is dominant in such an epoch,
possibly interacting with the cosmological constant.
Then \eqref{trace} becomes
\begin{equation}
    -3\dot{H} -6H^2 + 2\Lambda_0 + 3s_0^2(1 - 8\pi G g_3) = 0.
    \label{s0_domin}
\end{equation}
Hence $\ddot{a} > 0$ means that we need
\begin{equation}
    s_0^2(1 - 8\pi G g_3) > H^2 - \frac{2}{3}\Lambda_0.
    \label{accel_torsion}
\end{equation}
From \eqref{s0_domin}, for $g_3 \neq 1/(8\pi G)$,
we also find that 
\begin{equation}
    s_0^2 = \frac{1}{(1 -8\pi G g_3)}\left(\dot{H} + 2H^2 - \frac{2}{3}\Lambda_0\right),
    \label{tors_late}
\end{equation}
which, plugged into the first Friedmann equation
\eqref{Fried1} results in a differential equation for $H$:
\begin{multline}
    \frac{-3 \dot{H}+(\Lambda_0 - 3H^2)(1+8\pi G g_3)}{8 \pi  G(1-8\pi G g_3)}
    -g_3\frac{(2\Lambda_0 - 6H^2 -3\dot{H})}{1-8\pi G g_3}\\
    -3 g_1 \Bigg\{H^4-(H^2 + \dot{H})^2+
    \frac{1}{1-8\pi G g_3}\Big[H(4H\dot{H} + \ddot{H})\\
    -5H^2(2H^2 + \dot{H}-\frac{2}{3}\Lambda_0)\Big]
    + \frac{(2H^2 + \dot{H} - 2\Lambda_0/3)^2}{(
    1-8\pi G g_3)^2} \\
    + \frac{(4H\dot{H} + \ddot{H})^2}{4(
    1-8\pi G g_3)(2H^2 + \dot{H} -2\Lambda_0/3)}
    \Bigg\}=0.
    \label{diff_H}
\end{multline}
Obviously now we are not allowed to have 
\begin{equation}
    \dot{H}+2 H^2-\frac{2}{3} \Lambda_0 = 0.
\end{equation}
This is however 
equivalent to $s_0 = 0$ by \eqref{tors_late} and hence would
be a contradiction to our assumption.

In theory, the solution of \eqref{diff_H} governs the
time evolution of the scale factor and via \eqref{tors_late}
also the time evolution of the torsion $s_0$. But it is not 
easy to solve \eqref{diff_H} in general. Hence let us consider
a specific example.

\subsubsection{Exponential expansion}
One special case for accelerated expansion is 
the exponential ansatz
${a \propto \exp(Ct)}$ for some 
$C>0$ with $[C] =\SI{}{L}^{-1} $.
We then have that $H = C$ and $\dot{H} = 0$.
This is most commonly used in $\Lambda$CDM cosmology in
conjunction with the cosmological constant as the sole source
of dark energy.

In this approximation, equation \eqref{trace} becomes
\begin{equation}
    -6H^2 +2\Lambda_0 + 3s_0^2(1 - 8\pi G g_3) = 0
\end{equation}
at late times.
This is solved for $s_0$ as 
\begin{equation}
    s_0 = \pm \sqrt{\frac{2}{1 - 8\pi G g_3}\left(H^2 - \frac{1}{3}\Lambda_0\right)}.
    \label{s0_late}
\end{equation}
By a similar analysis as before, using that $H$ and $s_0$ are constant, Eq.~\eqref{Fried1} is reduced to 
\begin{equation}
    3g_1\left(-5H^2s_0^2 + s_0^4\right) + \frac{-3H^2 + \Lambda_0 + 3s_0^2(1 - 8\pi G g_3)}{8\pi G} = 0.
\end{equation}
With \eqref{s0_late} we then get
\begin{multline}
    \left(H^2-\frac{1}{3}\Lambda_0\right)\Bigg[
    -g_1\left(6H^2 + \frac{4}{3}\Lambda_0\right) +
    \frac{1}{8\pi G} \\
    - 2g_3\left(1 - 4\pi G g_3 - 40\pi G H^2 g_1\right)\Bigg] = 0.
\end{multline}
This equation is solved by either $H^2 = \Lambda_0/3$
or 
\begin{equation}
    H^2 = \frac{16 \pi G \left[-2 g_1 \Lambda_0+3 g_3 (4 \pi G g_3 -1)\right]+3}{48 \pi G g_1 (3 - 40 \pi G g_3)}
    \label{H2_val}
\end{equation}
with $g_3 \neq 3/(40\pi G)$.
If $H^2 = \Lambda_0/3$,
then it follows from \eqref{s0_late} that $s_0 = 0$, which 
contradicts our assumption.
The only remaining case is \eqref{H2_val}, giving  
\begin{equation}
    s_0 = \pm \frac{1}{2} \sqrt{\frac{8\pi G(10\Lambda_0 g_1 + 3g_3) -3}{6\pi G g_1(40\pi G g_3 - 3)}}.
\end{equation}
This case also ensures that the conservation equation \eqref{cons_lhs} is satisfied.
Applying all previously
stated approximations in this limit and using
\eqref{s0_late}, Eq.~\eqref{cons_lhs} becomes
\begin{multline}
    H \left(\Lambda_0-3H^2\right) \Big[3-16 \pi g_1 G \left(2 \Lambda_0+9 H^2\right) \\
    - 48\pi G g_3(1 - 4\pi G g_3 - 40\pi G H^2 g_1)\Big]=0.
\end{multline}
The l.h.s. of this equation indeed vanishes for $H^2$ as
in \eqref{H2_val}.

\medskip
Since the Universe already entered a phase of accelerated
expansion, the value for the Hubble parameter can be set
equal to the Hubble constant ${H_0 \approx 
\SI{70}{\kilo\metre\per\second\per\mega\parsec}}$
\cite{Planck_2018}.
For a given $g_3$, equation \eqref{H2_val} then reveals a 
simple inverse proportionality
between the parameter
$g_1$ and the cosmological constant $\Lambda_0$
\begin{equation}
    g_1 = \frac{-3 (1 - 8 \pi G g_3)^2}{16 \pi G\, [-2 \Lambda_0 + 3 H_0^2 (40 \pi  G g_3 - 3)]}.
    \label{g1_fct}
\end{equation}
Likewise are the behaviours of $g_2$ and $g_4$ as a function of 
$\Lambda_0$  obtained from Eq.~\eqref{parameters}:
\begin{equation}
    g_2 = \frac{-2\Lambda_0 + 3 H_0^2 (40 \pi G g_3-3)}{3 (1 - 8 \pi  G g_3)^2}
\end{equation}
\begin{equation}
    g_4 = \frac{-1}{8\pi G}\left[-\Lambda_0 + \frac{3 H_0^2 (3 - 40 \pi  G g_3)+2 \Lambda_0}{(1 - 8 \pi G g_3)^2}\right].
\end{equation}

\medskip
Let us consider the special circumstance where the expansion of the Universe is solely driven by torsion. In other words,
we set $\Lambda_0 = 0$ as is suggested by the
``Zero-Energy-Universe'' conjecture discussed in
\cite{Cosmol_const}. 
In that case the combination of the coupling constants $g_1$ and $g_2$ cancels the vacuum energy of matter, $g_4$. The trace equation \eqref{s0_late} then yields
\begin{equation}
    s_0 = \pm \sqrt{\frac{2}{1 - 8\pi G g_3}}\,H_0,
\end{equation}
whereas \eqref{H2_val} reduces to
\begin{equation}
     H_0^2 = \frac{48 \pi G g_3 (4 \pi G g_3 -1)+3}{48 \pi G g_1 (3-40 \pi G g_3)},
    \label{exp_H}
\end{equation}
which further means
\begin{equation}
    s_0 = \pm \frac{1}{2} \sqrt{\frac{1-8 \pi G g_3}{2\pi G g_1(3-40 \pi G g_3)}}.
    \label{exp_s0}
\end{equation}
Equations \eqref{exp_H} and \eqref{exp_s0} again
emphasize the relevance of the quadratic Riemann-Cartan term
via the appearance of $g_1$.

The parameter $g_3$ only ever appears multiplied by multiples of $M_p^{-2}$ and is subsequently
added to constants of order unity. The torsional
contribution coming from the tensor \eqref{quadr_tors} is thus only relevant if $g_3 \gtrsim M_p^2$.
For $\abs{g_3} \ll M_p^2$ the parameters $g_1$, $g_2$ and $g_4$ 
admit a very simple form.
From \eqref{g1_fct} we get
\begin{equation}
    g_1 = \frac{1}{48\pi G H_0^2} = \frac{
    M_p^2}{6H_0^2}.
\end{equation}
We hence find the expected value of $g_1$ to be
\begin{equation}
    g_1 \approx \SI{4.45e119}{}.
    \label{g1_val}
\end{equation}
For the parameters $g_2$ and $g_4$ we find
\begin{align*}
    g_2 &= \frac{-1}{16\pi G g_1} = -3 H_0^2 \\
    g_4 &= \frac{-3}{128\pi^2 G^2 g_1} = -9 M_p^2 H_0^2
    \numberthis
\end{align*}
and therefore with \eqref{g1_val} we obtain
\begin{align*}
    g_2 &\approx \SI{-6.69e-84}{\giga\electronvolt^{2}}\\
    g_4 &\approx \SI{-1.19e-46}{\giga\electronvolt^{4}}.
    \numberthis
    \label{g23_val}
\end{align*}
As the parameter $g_2$ is related to the Riemann curvature
tensor of the maximally symmetric spacetime via 
$\tensor{\hat{R}}{_{\eta\alpha\xi\beta}} = -g_2(
\tensor{g}{_{\eta\xi}}\tensor{g}{_{\alpha\beta}} - 
\tensor{g}{_{\eta\beta}}\tensor{g}{_{\alpha\xi}}
)$ \cite{CCGG_DE}, a positive (negative) $g_2$ therefore implies an AdS (dS)
geometry of the ground state of spacetime. Hence for 
$\abs{g_3} \ll M_p^2$ the ground state of spacetime admits
a dS geometry.

With the assumptions and restrictions applied here, an
exponential expansion in a torsion dominated epoch is thus
seen
to be possible even if $\Lambda_0 = 0$.  Requesting in
addition $\abs{g_3} \ll M_p^2$ and consistency with the observed Hubble constant $H_0$
yields for the parameters $g_i$  the specific values given in
Eqs.~\eqref{g1_val} and \eqref{g23_val}. (Notice that the
vacuum energy inferred is in the
$\SI{}{\milli\electronvolt}^4$ range, which is the order of
magnitude discussed in~\cite{Prat:2021xlz}.)

\begin{figure}
\includegraphics[width=8.6cm,height=5cm]{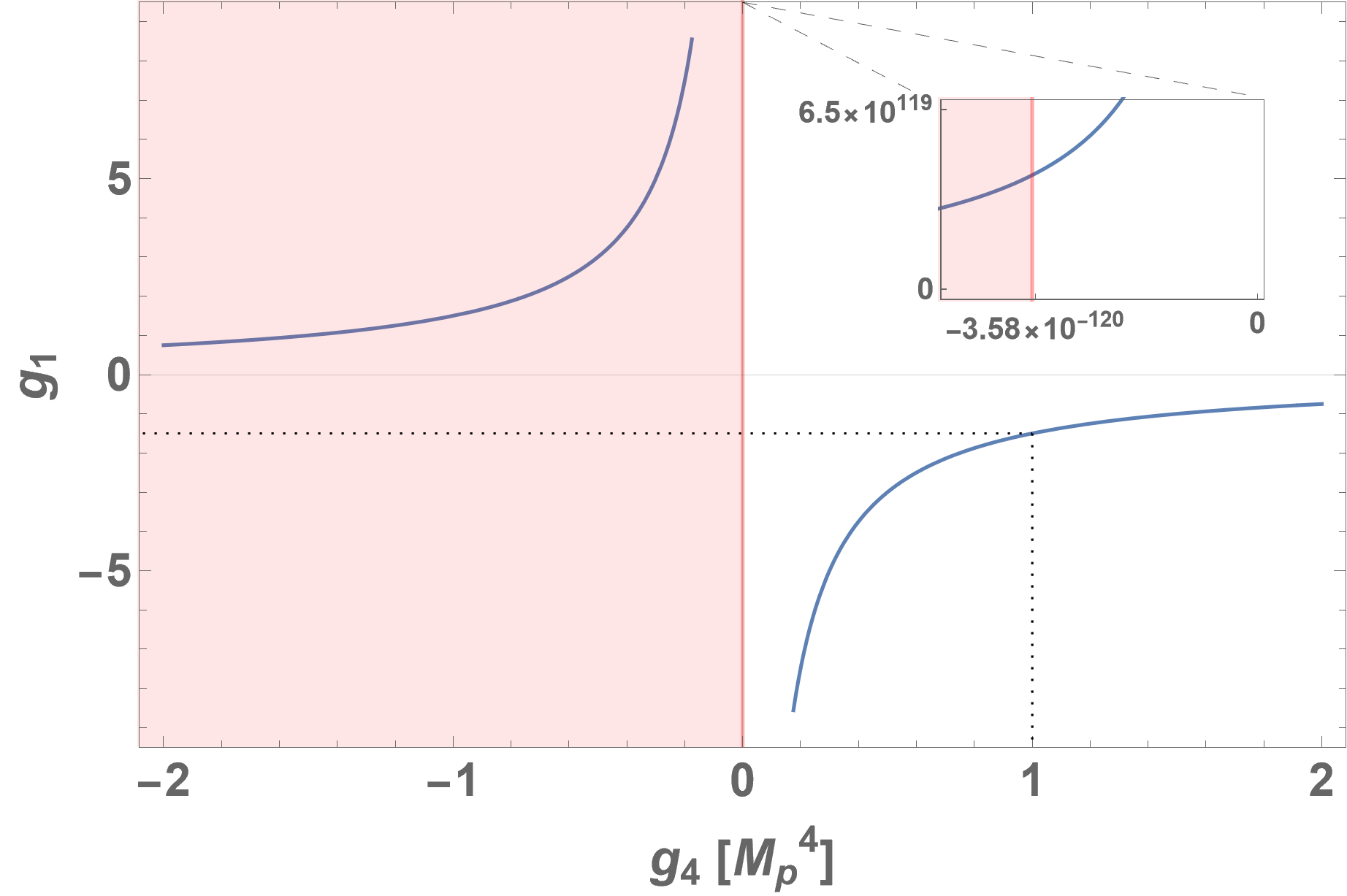}
\caption{Dependence of the parameter $g_1$ on the vacuum energy $g_4$ for the late times exponential expansion with
$\Lambda_0 = 0$. The red region indicates the forbidden values for $g_4$. The dotted line marks the naive cutoff
at Planck scales, for which $g_1 = -3/2$.}
\label{g1_g4}
\end{figure}

If $\abs{g_3} \gtrsim M_p^2$ the 
contribution from \eqref{quadr_tors} is relevant and leaves us
with an additional free parameter.
Here it is possible to express all parameters as functions of
the vacuum energy $g_4$:
\begin{align}
    g_1 &= \frac{-3 M_p^4}{2 g_4}\\
    g_2 &= \frac{g_4}{3 M_p^2}
\end{align}
and
\begin{equation}
    g_3 = M_p^2\left[\frac{15 M_p^2 H_0^2 \pm M_p H_0\sqrt{
3(75M_p^2 H_0^2 + 8g_4)}}{2 g_4} + 1\right].
\end{equation}
In particular we thus need $g_4 \geq -75 M_p^2 H_0^2/8 \sim 
\SI{-e-46}{\giga\electronvolt^{4}}$ and $g_4 \neq 0$.
The dependence of the parameter $g_1$ on the
vacuum energy $g_4$ is illustrated in Fig.\,\ref{g1_g4}.
Depending on the sign of $g_4$, the case $\abs{g_3} \gtrsim M_p^2$ also allows for an AdS geometry of the ground state
of spacetime.

\section{Conclusion}
\label{Concl}
For a specific ansatz of the free gravitational De Donder-Weyl Hamiltonian we investigated the corresponding CCGG equations which extend
Einstein's field equations by a trace-free Kretschmann
term and torsion contributions. 
After choosing a totally anti-symmetric, temporal ansatz for the torsion tensor, it was possible to align our theory with the Cosmological Principle and the energy density scaling behaviours of standard $\Lambda$CDM cosmology. 

\medskip
In a FLRW-Universe with perfect fluids accounting for the stress-energy density, the modified Friedmann equations were derived and given explicitly. 
Consistency has been ensured by verifying the conservation of the energy-momentum tensor of gravity on the l.h.s. of the CCGG
equations. 
In the Einstein-Cartan limit, with the Kretschmann
term set to zero, the modifications due to torsion turned out 
to be equivalent to spatial curvature contributions.
We thus argue that in such a case it would be possible to misinterpret the 
observationally inferred value 
of the spatial curvature $k$ 
in presence of torsion. The geometry type of the Universe
is not solely bound to the value of $k$ anymore but
depends also on the torsional parameter $s_0$.

\medskip
Finally, with the Kretschmann
term invoked, the modified Friedmann equations were investigated in early and late times, by making appropriate assumptions for the scale factor dependence.
In the RDE, we identified, in addition to the standard cosmology solution, a further, novel linear
time dependence of the scale factor by virtue of the interplay between radiation, the Kretschmann term and torsion. Such a linear dependence has already appeared in
Jordan-Brans-Dicke cosmology \cite{JBD} and has further
been shown to be in agreement with data in the 
so-called Power Law Universes \cite{Pow_law}. Linear Coasting
cosmologies
appear very alluring in this regard as they
do not suffer from the horizon problem nor the flatness 
problem while complying with data from SNeIa, gravitational lensing and high-redshift galaxies \cite{Lin_coast1,Lin_coast2}.

Moreover, we learned that in the late epoch torsion can account for a phase of accelerated expansion even without the presence of a 
cosmological constant.
For a specific case, namely exponential expansion, 
as expected in the  dark energy era, the values of the
parameters $g_1$,
$g_2$ and $g_4$ were derived from the observed value of the current late-time Hubble constant
in the case that ${\abs{g_3} \ll M_p^2}$. For the vacuum energy of matter, $g_4$, we find a value in the $\SI{}{\milli\electronvolt}^4$ range.
We therefore claim that our model is able to provide a 
geometrical 
explanation for the late time accelerated expansion of 
the Universe through torsion 
without resorting to the introduction of a cosmological 
constant.

For $\abs{g_3} \gtrsim M_p^2$ we were additionally able to
provide a lower bound on the vacuum energy of matter. We have
yet to investigate the implications of this finding.

\medskip
Work in this area is still in progress. A numerical evaluation of the modified Friedmann equations as shown in this paper will be presented in the near future. 
That numerical 
analysis is not restricted to the simplified monomial ansatz
of the scale factor in the RDE which we used in this paper.
Indeed polynomial behaviour is seen for certain solutions
which allows for more freedom in the exploration of the 
expansion history of the Universe. Early Dark Energy (EDE)
solutions appear as possible candidates in this regard \cite{EDE1,EDE2,EDE3}.
Ultimately, in order to check the consistency of our ansatz with the full set of cosmological observations, we envisage to carry out an MCMC analysis for the late data, but also work out the perturbation theory in detail to include the CMB data. 

\medskip
Ideas to be followed beyond that certainly include modifying the free gravitational Hamiltonian 
and the models of the torsion tensor in compliance with the  Cosmological Principle \cite{Iosifidis1,Iosifidis2}.
Last but not least: Discovering an independent way of measuring the parameter $g_1$ is one of the key pending tasks for the future.

\section*{Acknowledgements}
The authors are indebted to the ``Walter Greiner-Gesell\-schaft zur F\"{o}rderung
der physi\-ka\-lischen Grundlagenforschung e.V.'' (WGG) in Frankfurt for their support.\,
AvdV, DV and JK especially thank the Fueck Stiftung for support. \\
The authors also wish to thank David Benisty, Eduardo Guendelman, Horst Stöcker, Yilin Cheng, Wei Li and Peter Hess for valuable discussions.

 \bibliography{biblio}

\end{document}